\documentclass[conference,a4paper]{APSIPA2025}
\usepackage{amsmath}
\usepackage{graphicx}
\usepackage{multirow}
\usepackage{threeparttable}
\usepackage[backend=biber,style=ieee,]{biblatex}
\usepackage{amsmath}  
\usepackage{amssymb}  
\usepackage{bm}  
\usepackage{pifont} 
\usepackage{adjustbox}  
\usepackage{graphicx}   
\usepackage{geometry}
\usepackage{subcaption} 
\usepackage{fancyhdr}

\fancypagestyle{firststyle}{
  \fancyhf{}
  \fancyhead[C]{2025 Asia Pacific Signal and Information Processing Association Annual Summit and Conference (APSIPA ASC)}
}

\begin{document}

\title{A High-Quality and Low-Complexity Streamable Neural Speech Codec with Knowledge Distillation}
\author{
\authorblockN{En-Wei Zhang$^*$, Hui-Peng Du$^*$, Xiao-Hang Jiang, Yang Ai$^\dagger$, and Zhen-Hua Ling
}

\authorblockA{
National Engineering Research Center of Speech and Language Information
Processing, \\University of Science and Technology of China, Hefei, P. R. China\\
E-mail: \{zhangenwei, redmist, jiang\_xiaohang\}@mail.ustc.edu.cn, yangai@ustc.edu.cn, zhling@ustc.edu.cn}
}

\maketitle
\renewcommand{\thefootnote}{}
\footnotetext{$*$ Equal contribution.}
\renewcommand{\thefootnote}{\arabic{footnote}}
\thispagestyle{firststyle}
\pagestyle{fancy}

\begin{abstract}
While many current neural speech codecs achieve impressive reconstructed speech quality, they often neglect latency and complexity considerations, limiting their practical deployment in downstream tasks such as real-time speech communication and efficient speech compression.
In our previous work, we proposed StreamCodec, which enables streamable speech coding by leveraging model causalization and a scalar-vector-combined quantization strategy, but its reconstructed quality and complexity still have room for improvement. 
Therefore, this paper proposes an improved iteration of StreamCodec, named StreamCodec2. 
The StreamCodec2 supports streamable and lightweight speech coding by adopting a fully causal architecture and reducing the convolutional channels. 
To compensate for the speech quality degradation caused by model causalization and pruning, we introduce a non-causal, high-complexity teacher codec to guide the training of StreamCodec2 through knowledge distillation.
Experimental results demonstrate that our proposed StreamCodec2, trained with the knowledge distillation strategy, can achieve high-quality speech reconstruction while maintaining low latency (only 20 ms), low computational complexity (only 910 MFLOPs), and low model complexity (only 5.4 M parameters).
\end{abstract}

\section{Introduction}
\renewcommand{\thefootnote}{}
\footnote{$^\dagger$ Corresponding author. This work was funded by the Anhui Province Major Science and Technology Research Project under Grant S2023Z20004, the National Nature Science Foundation of China under Grant 62301521 and the Anhui Provincial Natural Science Foundation under Grant 2308085QF200.}
 \renewcommand{\thefootnote}{\arabic{footnote}}
\addtocounter{footnote}{-1}
A speech codec can encode input speech into discrete representations and subsequently reconstruct the speech from these discrete representations.
It has been applied in many downstream tasks, e.g., speech communication \cite{salami1994toll, du2024apcodec+}, text-to-speech (TTS) \cite{zhang2023speechtokenizer,wang2023neural,borsos2023audiolm}, and speech enhancement \cite{xue2024low}. 
Taking speech communication as an example, a speech codec enables efficient transmission of speech signals over networks by compressing the speech into a compact form and reconstructing it at the receiver with acceptable quality.

\begin{figure*}
    \centering
    \includegraphics[width=0.84\linewidth]{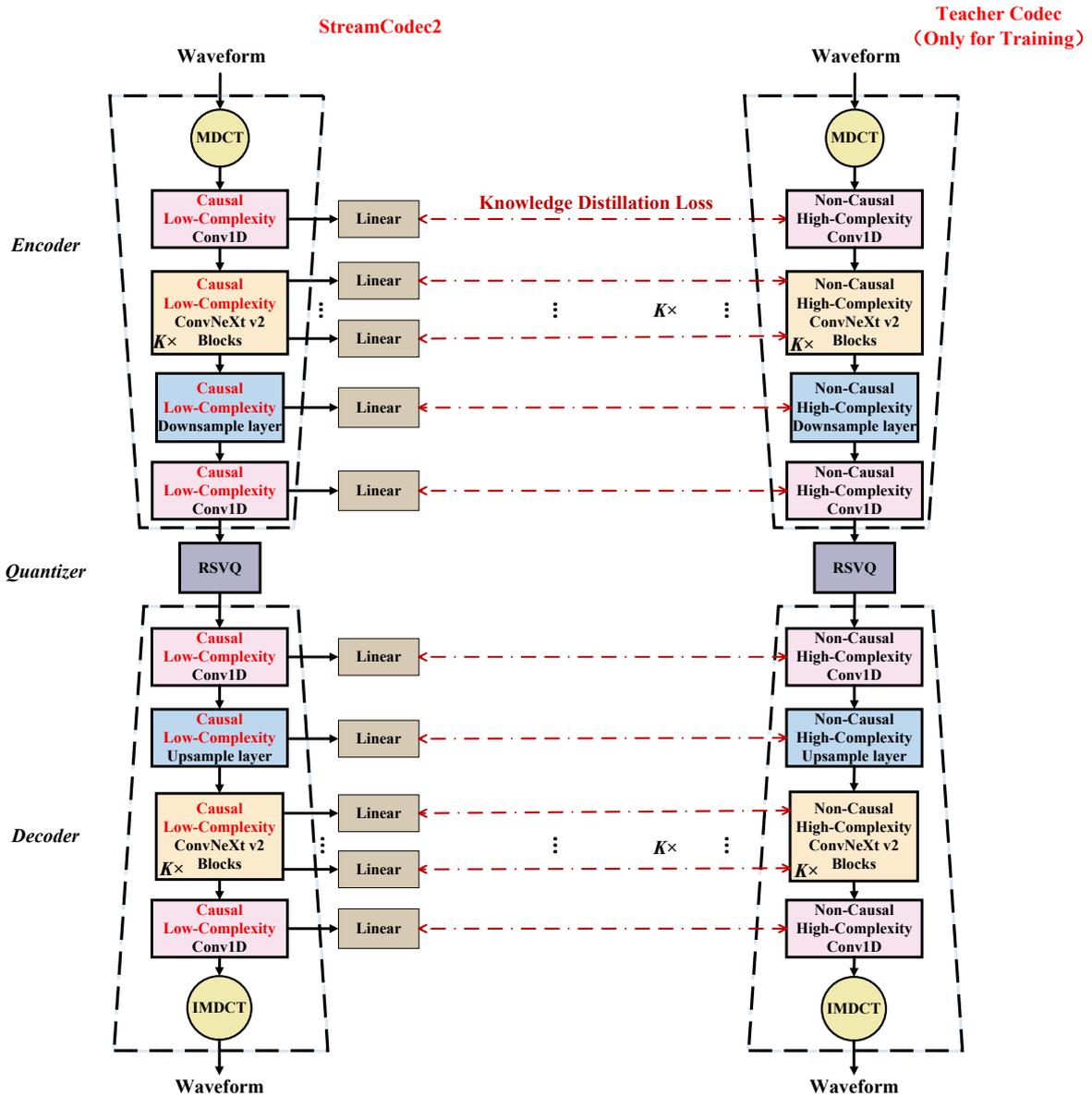}
    \caption{An overview of the model structure and knowledge distillation strategy of the proposed StreamCodec2.}
    \label{fig:1}
\end{figure*}
Speech codecs generally consist of three components, i.e., an encoder, a quantizer, and a decoder. 
The encoder transforms input speech or speech features into frame-level continuous representations, which are then discretized by the quantizer into compact discrete representations. 
Finally, the decoder reconstructs the speech waveform from these discrete representations. 
Reconstruction quality, bitrate, complexity, and latency are key metrics for evaluating the performance of a speech codec. 
Among them, reconstruction quality reflects the codec's ability to faithfully recover speech signals, while bitrate represents the coding efficiency. 
Complexity, which can be divided into computational complexity and model complexity, determines the feasibility of deploying the codec on resource-constrained devices such as embedded systems. 
Latency refers to the initial time required for the codec to start processing, which is particularly critical in real-time speech communication scenarios. 
Excessive latency can lead to noticeable conversational delays, degrading the user experience in interactive communication.

Early speech codecs such as Opus \cite{valin2013high} and EVS \cite{dietz2015overview} used digital signal processing (DSP) algorithms to encode speech into discrete representations without the need for data training. 
However, DSP-based methods often suffer from poor reconstruction quality under low-bitrate conditions.
Recently, SoundStream \cite{zeghidour2021soundstream} and Encodec \cite{defossez2023high} introduced neural networks to speech coding tasks, leveraging causal convolutional layers and adversarial training to enable high-quality speech reconstruction with low model latency. 
While their speech quality surpasses that of DSP-based methods, their performance still remains unsatisfactory.
In pursuit of higher speech quality, some subsequent works adopt more sophisticated network structures, e.g., DAC \cite{kumar2024high}, while neglecting the control of computational cost and latency. This makes them difficult to deploy in practical applications.
Our previous work, APCodec \cite{ai2024apcodec}, proposed a dual-path structure for amplitude and phase coding. 
Compared to waveform-coding-based methods, APCodec achieves further complexity reduction while preserving the same level of speech quality. 
Building on this, we further proposed StreamCodec \cite{jiang2025streamable}, which adopts a simpler and more compact modified discrete cosine transform (MDCT) spectrum as the coding target and introduces a new scalar-vector-combined quantization strategy. 
It explicitly considers model latency, making it suitable for real-time applications. 
However, its speech reconstruction quality and complexity still have room for further improvement.

To address these limitations, this paper proposes an improved version of StreamCodec \cite{jiang2025streamable}, named StreamCodec2, which introduces a knowledge distillation training strategy to enable higher-quality and lower-complexity efficient speech coding.
The StreamCodec2 encodes, quantizes, and decodes the MDCT spectrum of the input speech, and finally reconstructs the speech waveform through inverse MDCT (IMDCT). 
To support low-latency streamable generation, StreamCodec2 employs a fully causal network architecture for both the encoder and decoder, which are primarily composed of 1D causal convolution layers and causal ConvNeXt v2 blocks. 
To achieve low-complexity coding, StreamCodec2 further reduces the number of channels in convolution operations compared to StreamCodec. 
Model causalization and pruning inevitably lead to degradation in speech quality. 
To mitigate this, StreamCodec2 introduces a knowledge distillation training strategy \cite{hinton2015distilling} that forces it to approach the performance of a non-causal, high-complexity teacher codec, enabling high-quality speech coding. 
In our experiments, we discuss and analyze several distillation strategies and validate the effectiveness of the proposed knowledge distillation method. Through distillation learning, StreamCodec2 is able to achieve high-quality speech reconstruction while maintaining the advantages of low latency and low complexity.

The rest of this paper is organized as follows. 
We describe the details of the proposed StreamCodec2 in Section \ref{sec:pm} and show the experimental setups and results in Section \ref{sec:es} and \ref{sec:ra}, respectively. 
Finally, we give conclusion in Section \ref{sec:c}.

\section{Proposed Method}
\label{sec:pm}
\subsection{Overview}
An overview of the proposed StreamCodec2 is shown in Fig. \ref{fig:1}. 
The StreamCodec2 consists of an encoder, a quantizer, and a decoder. 
The encoder encodes the MDCT spectrum of the input speech, and the quantizer discretizes the encoded features into discrete tokens. 
The decoder then recovers the MDCT spectrum from the quantized results and reconstructs the speech waveform through IMDCT.
To compensate for the degradation in reconstructed speech quality caused by model causalization and pruning, we adopt a knowledge distillation training strategy and select non-causal, high-complexity codec as teacher model to guide the learning of StreamCodec2. 
In the following sections, we provide a detailed introduction to the model structure and training criteria of StreamCodec2. 

\begin{figure}[t]
    \centering
    \includegraphics[width=0.95\linewidth]{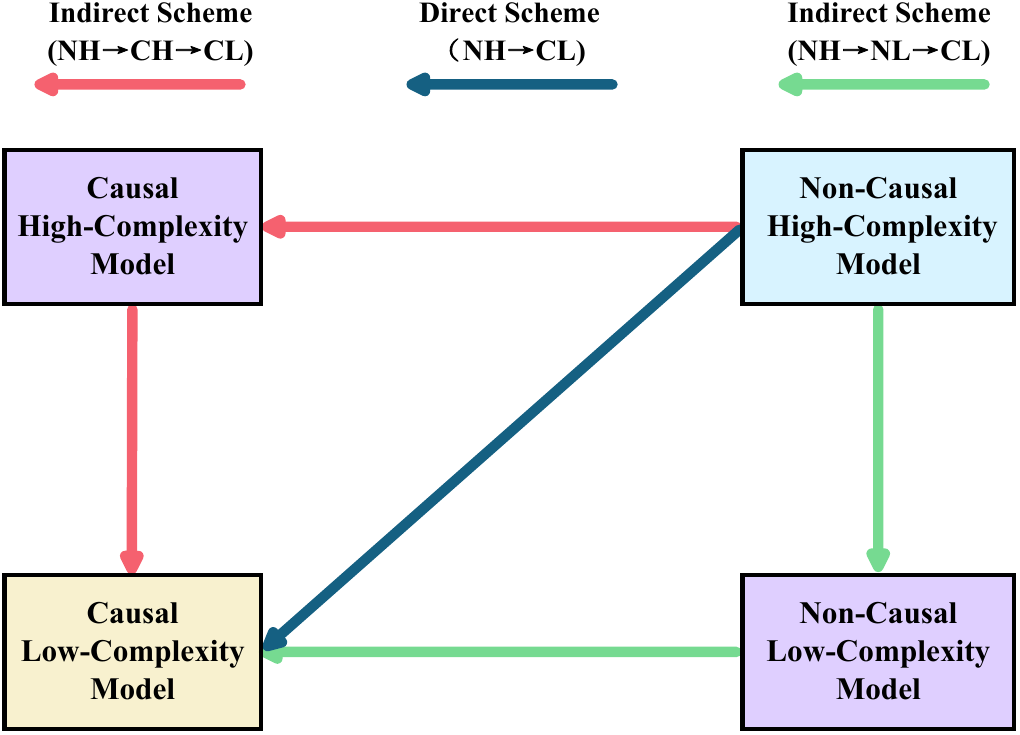}
    \caption{Illustration of the three distillation schemes for StreamCodec2.}
    \label{fig:2}
\end{figure}


\subsection{Model Structure}

The encoder and decoder of StreamCodec2 adopt a mirror-symmetric structure based on fully causal convolutions. 
Compared to StreamCodec, it uses fewer convolutional channels to reduce complexity.
Taking the encoder as an example, it takes the MDCT spectrum of the input waveform as input, first processes it through a causal input convolution layer, and then extracts features using $K$ causal ConvNeXt v2 blocks. 
The causal ConvNeXt v2 block, adapted from the original ConvNeXt v2 block \cite{woo2023convnext} to support causality, uses a 7$\times$1 causal convolutional layer with a channel of $C_L$ to extract preliminary features. 
It then employs linear layers to project the features to an intermediate dimension $C_H$ to capture richer information before compressing them back to dimension $C_L$, where $C_L<C_H$.
These features are then downsampled using a causal downsampling convolution to reduce the frame rate, facilitating the discretization into lower-bitrate discrete tokens.
Finally, another causal convolutional layer processes the output of the causal ConvNeXt v2 block to generate the encoded features. 
The structure of the decoder mirrors that of the encoder, except that the causal downsampling convolution is replaced by a causal upsampling convolution.

The quantizer in StreamCodec2 adopts a residual scalar-vector quantization (RSVQ) strategy, consisting of a scalar quantizer (SQ) \cite{mentzer2024finite, strom2022squashed} and two improved vector quantizers (IVQs) \cite{zheng2025ervq}, connected in a residual manner. 
The SQ first generates coarse-grained quantized features, while IVQs, compared to conventional VQs, adopt a codebook clustering strategy during training to further refine these features, obtain fine-grained quantized representations, and improve codebook utilization.

\subsection{Knowledge-Distillation-based Training Criteria}
\label{ssec:pp}
To compensate for the reduced learning capacity caused by model causalization (i.e., the use of causal convolutions) and pruning (i.e., reducing the number of channels), we adopt a knowledge distillation strategy for StreamCodec2.
As illustrated in Fig. \ref{fig:1}, we pre-train a non-causal, high-complexity speech codec with a structure similar to StreamCodec2 as the teacher model. It leverages the intermediate features generated from the input speech to guide the learning of the corresponding intermediate features in StreamCodec2.
For StreamCodec2, we sequentially define the features output by each module from the encoder to the decoder (see Fig. \ref{fig:1}) as $\hat{\bm{O}}_{1}\in\mathbb R^{S_1\times F_1}, \dots, \hat{\bm{O}}_{n}\in\mathbb R^{S_n\times F_n},..., \hat{\bm{O}}_{N}\in\mathbb R^{S_N\times F_N}$, where $N$ represents the total number of layers for distillation. 
$S_n$ and $F_n$ denote the dimension and the number of frames of $\hat{\bm{O}}_{n}$, respectively. 
In our implementation, $N = 2k + 6$. 
The outputs of the teacher codec at corresponding positions are respectively defined as $\widetilde{\bm{O}}_{1}\in\mathbb R^{T_1\times F_1}, \dots, \widetilde{\bm{O}}_{n}\in\mathbb R^{T_n\times F_n},..., \widetilde{\bm{O}}_{N}\in\mathbb R^{T_N\times F_N}$, where $T_n$ denotes the number of frames of $\widetilde{\bm{O}}_{n}$. 
To address the dimensional mismatch between the corresponding intermediate features of the teacher and student models, we introduce trainable projection linear layer with weight $\bm{W}_n\in\mathbb R^{T_n\times S_n}$ to align the dimensions of $\hat{\bm{O}}_{n}$ and $\widetilde{\bm{O}}_{n}$.
Consequently, the knowledge distillation loss can be formulated as:
\begin{align}
    \mathcal{L}_{KD} &=\frac{1}{N}\sum_{n=1}^{N}\mathbb{E}_{(\hat{\bm{O}}_n,\widetilde{\bm{O}}_n)}\left\| \bm{W}_n\hat{\bm{O}}_n - \widetilde{\bm{O}}_n \right\|_2.
\end{align}

Besides, we also maintain the adversarial loss $\mathcal{L}_{adv}$, the feature matching loss $\mathcal{L}_{FM}$, the MDCT spectrum loss $\mathcal{L}_{MDCT}$, the mel spectrogram loss $\mathcal{L}_{Mel}$, the codebook loss $\mathcal{L}_{cb}$, and the commitment loss $\mathcal{L}_{com}$, defined in \cite{jiang2024mdctcodec}. 
The overall loss for training StreamCodec2 is defined as:
\begin{align}
\mathcal{L} &= \mathcal{L}_{adv} + \mathcal{L}_{FM} + \lambda_{MDCT}\mathcal{L}_{MDCT} + \nonumber \\
&\quad \lambda_{Mel}\mathcal{L}_{Mel} + \lambda_{cb}\mathcal{L}_{cb} + \lambda_{com}\mathcal{L}_{com} + \lambda_{KD}\mathcal{L}_{KD},
\end{align}
where $\lambda_{MDCT}, \lambda_{Mel}, \lambda_{cb}, \lambda_{com}$ and $\lambda_{KD}$ are hyperparameters.

\begin{table*}[ht]
\renewcommand{\arraystretch}{1.1}
	\centering
	\caption{Objective experimental results of the teacher model, student model, and three distilled models using different distillation schemes of the proposed StreamCodec2 on the LibriTTS test set. The p-values indicate the significance of the paired t-tests between StreamCodec2 ($NH\to CL$) and the student model.}\label{tab1}
		\begin{tabular}{c |c c c c| c c}
			\hline
			\hline
			 & {LSD} & {STOI} & {PESQ} & {ViSQOL} & {FLOPs} & {Param.}\\
			\hline
            {Teacher Model} & 0.837 & 0.946 & 3.132 & 4.463 & 1237 M & 7.2 M\\
      \hline

            {Student Model} & 0.887 & 0.928 & 2.650 & 4.290 & \multirow{5}{*}{\textbf{910 M}} & \multirow{5}{*}{\textbf{5.4 M}} \\
            \cline{1-5}

\multirow{2}{*}{StreamCodec2 ($NH\to CL$)} & \textbf{0.869} & \textbf{0.933} & \textbf{2.744} & \textbf{4.313} &  &  \\
& \textbf{($\bm{p< 0.01}$)} & \textbf{($\bm{p< 0.01}$)} & \textbf{($\bm{p< 0.01}$)} & \textbf{($\bm{p< 0.01}$)} & & \\
\cline{1-5}
{StreamCodec2 ($NH \to CH \to CL$)} & 0.870 & 0.930 & 2.673 & 4.294 &  & \\ 
\cline{1-5}
{StreamCodec2 ($NH \to NL \to CL$)} & 0.874 & 0.931 & 2.741 & 4.305 &  & \\

			\hline
			\hline
	\end{tabular}
\end{table*}

\section{Experiments Setups}
\label{sec:es}
\subsection{Dataset}
Experiments were conducted on the LibriTTS speech dataset \cite{zen2019libritts}. 
This dataset contains 585 hours of speech recordings from 2,500 different speakers. 
We selected 149,708 utterances as the training set and built a test set containing 4,936 utterances. 
All speech utterances in the dataset were resampled to 16 kHz for our experiments. 
\subsection{Comparison Models}
In our experiments, we compared the student model, teacher model and models trained with three different distillation strategies (see Fig. \ref{fig:2}) of StreamCodec2. 
The details are as follows.

\begin{itemize}[leftmargin=*]

\item {}{\textbf{Student Model}:}
StreamCodec2 trained without the knowledge distillation strategy, which can be regarded as the lower bound of StreamCodec2's performance. 
In terms of its model structure, we set $C_L=200$ and $C_H=400$ for ConvNeXt v2 blocks.
    
\item {}{\textbf{Teacher Model}:}
A non-causal, high-complexity version of StreamCodec2 trained without causality and channel reduction, serving as the upper bound for StreamCodec2's performance and providing guidance for knowledge distillation. 
In terms of its model structure, we set $C_L=256$ and $C_H=512$ for ConvNeXt v2 blocks.

\item {}{\textbf{StreamCodec2 ($\bm{NH\to CL}$)}:}
This model adopts the direct distillation scheme (see blue path in Fig. \ref{fig:2}), where the non-causal high-complexity model serves as the teacher to directly distill the causal low-complexity student model. 

\item {}{\textbf{StreamCodec2 ($\bm{NH \to CH \to CL}$)}:}
This model adopts the indirect distillation scheme (see red path in Fig. \ref{fig:2}) in two steps: first, the non-causal high-complexity model distills the causal high-complexity model; then, the distilled causal high-complexity model is further used to distill the causal low-complexity student model. 

\item {}{\textbf{StreamCodec2 ($\bm{NH \to NL \to CL}$)}:}
This model also adopts a two-step indirect distillation scheme (see green path in Fig. \ref{fig:2}), where the non-causal high-complexity model first distills the non-causal low-complexity model, which is then used to distill the causal low-complexity student model.
\end{itemize}

All the above models use $K=8$ ConvNeXt blocks and are evaluated under a fixed bitrate of 1.7 kbps to ensure fair comparisons. 
Their specific model settings and loss function hyperparameters $\lambda_{MDCT}, \lambda_{Mel}, \lambda_{cb}$ and $\lambda_{com}$ are directly borrowed from \cite{jiang2025streamable}. 
For the three distillation models, the distillation weight is set to $\lambda_{KD}=0.01$. 
Except for the teacher model, all other models maintain a fixed latency of 20 ms.
All models are trained for 900k steps using the AdamW optimizer \cite{kingma2014adam} with an initial learning rate of 0.0002.


\subsection{Evaluation Metrics}

To assess the influence of different distillation approaches on reconstructed speech quality, we employ multiple objective evaluation metrics, including log-spectral distance (LSD), which measures the spectral differences between speech signals; short-time objective intelligibility (STOI) \cite{taal2010short}, designed to evaluate speech intelligibility; perceptual evaluation of speech quality (PESQ), which gauges perceived speech quality by comparing degraded and reference speech signals; and virtual speech quality objective listener (ViSQOL) \cite{chinen2020visqol}, which assesses speech quality in interactive contexts.
To evaluate computational and model complexity, we also calculate the floating point operations (FLOPs) \cite{mcmahon1986livermore} and the number of model parameters (Param.) for each model.

\section{Results and Analysis}
\label{sec:ra}

\subsection{Main Experimental Results}
The objective experimental results of the teacher model, student model, and three distilled models using different distillation schemes of the proposed StreamCodec2 are shown in Table \ref{tab1}. 
By comparing the teacher model and the student model, we observe that although model causalization and pruning lead to improvements in both computational complexity and model complexity, the reconstructed speech quality exhibits a significant decline across all objective quality metrics.
This degradation in performance makes it difficult to meet the speech quality requirements of downstream tasks, e.g., speech communication, limiting its practical applicability. 

However, as shown in Table \ref{tab1}, after applying our proposed distillation strategies, all three distillation methods achieve noticeable improvements in speech quality metrics compared to the model without distillation (i.e., the student model). 
This demonstrates that the introduction of knowledge distillation, regardless of the specific method used, can effectively enhance the reconstructed speech quality.
For StreamCodec2 ($NH\to CL$), which adopts the direct distillation scheme, we conducted paired t-tests on the speech quality evaluation metrics against the student model to determine whether the differences between them were statistically significant. The results show that all p-values are below 0.01, indicating that the introduction of the direct distillation strategy leads to a significant improvement in StreamCodec2 over the student model.
To investigate the effects of different distillation methods, Table \ref{tab1} also presents a comparison of the speech reconstructed by the StreamCodec2 models under the three distillation strategies. 
The results show that the direct distillation method outperforms the other two indirect strategies across all speech quality metrics. 
This may be because multi-stage distillation can lead to a loss of useful information or insufficient transfer of fine-grained representations, reducing the effectiveness of knowledge distillation compared to a direct approach. 
These findings indicate that for this specific task, adopting direct distillation is preferable to staged multi-step distillation, as it simplifies the training process while improving speech reconstruction quality.
Since the knowledge distillation strategy affects only the model training rather than inference, it does not increase the computational or model complexity. 
Therefore, this strategy is well-suited for practical applications, as it enhances the quality of reconstructed speech without introducing additional computational or storage overhead. 
With the aid of knowledge distillation, the proposed StreamCodec2 is able to maintain high-quality speech coding while requiring only 910 MFLOPs of computational cost and 5.4 M model parameters.

\begin{figure}[t!]
    \centering
    \includegraphics[width=0.95\linewidth]{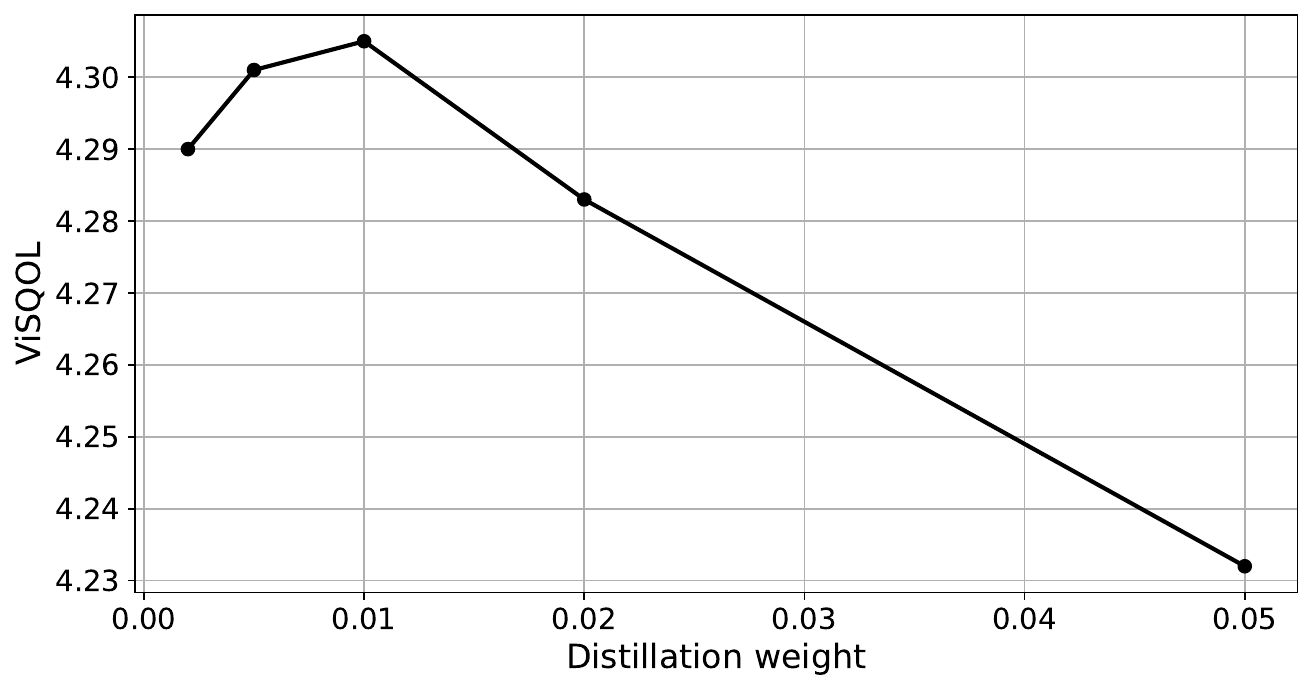}
    \caption{Variation of ViSQOL scores of reconstructed speech from StreamCodec2 ($NH\to CL$) with different knowledge distillation loss weights $\mathcal L_{KD}$.}
    \label{fig:3}
\end{figure}


\subsection{Analysis and Discussion}
In this section, we analyze the impact of distillation weights and distillation positions on the performance of StreamCodec2. 
Based on the results from the previous section, all experiments in this section are conducted using StreamCodec2 ($NH\to CL$).

\subsubsection{Analysis on Distillation Weight}
To analyze the impact of the knowledge distillation loss weight on the reconstructed speech quality of StreamCodec2, we conducted a series of experiments with different values of $\mathcal L_{KD}$.
Specifically, we set $\mathcal L_{KD}$ to 0.002, 0.005, 0.01, 0.02, and 0.05, respectively. 
For simplicity, we only calculated the ViSQOL values, and the curve showing their variation with different distillation weights $\mathcal L_{KD}$ is presented in Fig. \ref{fig:3}. 
It can be observed that as the distillation weight increases from a small value, the ViSQOL score of the reconstructed speech improves, indicating that the knowledge distillation strategy gradually becomes effective. 
The ViSQOL score reaches its peak when the weight is set to 0.01, after which the quality begins to degrade as the weight continues to increase. 
This may be because an excessively large distillation weight shifts the model’s learning objective from reconstructing the speech itself to overfitting the teacher's outputs. 
These findings indicate that the distillation weight should be carefully selected to balance the benefits of knowledge distillation with the preservation of the reconstruction objective, ensuring it is neither too small nor too large in practical deployment.

\subsubsection{Analysis on Distillation Position}
To investigate the impact of distillation at different locations, we conducted experiments by ablating the distillation positions.
Specifically, we removed the distillation losses from the output end of the up/downsampling layers (denoted as w/o UP/DO) and input/output convolutions (denoted as w/o I/O), respectively. 
The ViSQOL results of the speech quality generated by these models are shown in Table \ref{tab2}. 
It can be observed that applying knowledge distillation across all modules yields better speech quality than models where distillation is omitted at certain positions. 
This indicates that in knowledge distillation, utilizing sufficient and comprehensive distillation positions allows the teacher model to provide more effective guidance, leading to improved performance.

\begin{table}[t!]
\renewcommand{\arraystretch}{1.1}
	\centering
	\caption{ViSQOL scores of reconstructed speech from StreamCodec2 ($NH\to CL$) under ablation of knowledge distillation losses at different positions.}\label{tab2}
		\begin{tabular}{c c c c}
			\hline
			\hline
		   &{ w/o UP/DO}&{ w/o  I/O}&{StreamCodec2}\\
                \hline
             {ViSQOL}&4.296&4.283&4.313\\
			
			\hline
			\hline
	\end{tabular}
\end{table}

\section{Conclusion}
\label{sec:c}
This paper proposes StreamCodec2, an improved iteration of our previously developed StreamCodec, enabling streamable and lightweight neural speech coding through the adoption of a fully causal architecture and reduced convolutional channels. 
To address the speech quality degradation caused by model causalization and pruning, we introduce a non-causal, high-complexity teacher codec to guide the training of StreamCodec2 through knowledge distillation. 
Experimental results demonstrate that StreamCodec2 trained with this strategy can achieve high-quality speech reconstruction while maintaining low latency and low complexity.
In future work, we will focus on further improving the reconstructed speech quality and conducting experiments on additional audio datasets to evaluate the generalization of StreamCodec2.

\printbibliography

\end{document}